\makeatletter
\declare@file@substitution{revtex4-1.cls}{revtex4-2.cls}
\makeatother
\documentclass[onecolumn,onecolappendix]{aastex631}

\usepackage{bm}
\usepackage{amstext}
\usepackage{epstopdf}

\defcitealias{solomon:87}{S87}

\graphicspath{{./}{figures/}}
\begin{document}

\title{Rotation of Polarization Angle in Gamma-Ray Burst Prompt Phase$-$\uppercase\expandafter{\romannumeral2}. The Influence of The Parameters}

\author{Jia-Sheng Li}
\altaffiliation{These authors contributed equally to this work.}
\affiliation{Center for Theoretical Physics and College of Physics, Jilin University, Changchun, 130012, China; lanmixiang@jlu.edu.cn \\}
\author{Hao-Bing Wang}
\altaffiliation{These authors contributed equally to this work.}
\affiliation{Center for Theoretical Physics and College of Physics, Jilin University, Changchun, 130012, China; lanmixiang@jlu.edu.cn \\}
\author[0000-0001-5641-2598]{Mi-Xiang Lan}
\affiliation{Center for Theoretical Physics and College of Physics, Jilin University, Changchun, 130012, China; lanmixiang@jlu.edu.cn \\}

\begin{abstract}

In addition to the light curve and energy spectrum, polarization is also important for inferring the physical properties of the Gamma-ray burst (GRB). Rotation of the polarization angle (PA) with time will cause depolarization of the time-integrated polarization degree. However, it is rarely studied before. Here, we use a magnetic reconnection model with a large-scale ordered aligned magnetic field in the emitting region to study the influence of the parameters on the PA rotations in GRB prompt phase. We find that half-opening angle of the jet $\theta_{j}$, the observational angle $\theta_{V}$, and the bulk Lorentz factor $\Gamma$ all have significant impacts on the PA rotations. The PA rotations are affected by the product value of $\theta_{j}\Gamma_{0}$ ($\Gamma_{0}$ is the normalization factor of $\Gamma$ with $\Gamma(r)=\Gamma_{0}(r/r_{0})^{s}$), but are roughly independent of the concrete values of both $\theta_{j}$ and $\Gamma_{0}$. For the typical parameters, the changes of the PA within $T_{90}$ ($\triangle$PA) would be within ($12^\circ$, $66^\circ$) for slight off-axis observations, where $T_{90}$ is the duration of the burst with the accumulated flux density ranging from $5\%$ to $95\%$. The $q$ range for $\triangle$PA$>10^{\circ}$ becomes smaller with the increase of the product value of $\theta_{j}\Gamma_{0}$. The most significant PA rotation with $\triangle$PA$\sim90^{\circ}$ will happen when $\theta_{j}\Gamma_{0}>50$ and $1.0<q\leq1.2$.


\end{abstract}

\section{Introduction}\label{intro}

Gamma-ray bursts (GRBs) are the violent bursts of high-energy electromagnetic radiation in the universe. With a duration of 2 seconds as the boundary, GRBs can be divided into long and short bursts. It is generally believed that long bursts are generated by the collapse of massive stellar cores \citep{Mazzali2003,Woosley1993,Bloom1999,MacFadyen2001,Hjorth2003}. Short bursts are related to the mergers of the two neutron stars (NSs) or a NS with a black hole (BH) \citep{Narayan1992,Abbott2017,Goldstein2017,Lazzati2018}. Most of the observed GRB spectrum can be well fitted by the empirical formula Band function \citep{band93}, which is a broken power law, with low- and high-energy segments smoothly connected at $(\alpha_B-\beta_B)E_{peak}/(2+\alpha_B)$. $\alpha_{B}$ and $\beta_{B}$ represent the low- and the high-energy spectral indices of the photon number flux, respectively. And $E_{peak}$ is the peak energy of the $\nu F_\nu$ spectrum.

In order to explain the observations of GRB prompt phase, many models had been proposed, such as internal shock models \citep{Paczynski1994,Ress1994,Kobayashi1997,Sari1997,Daigne1998} and photosphere models \citep{Thompson1994,Eichler2000,Lundman2013}. \cite{Zhang2011} proposed the magnetic reconnection model, which assumes that the central engine of the GRB is highly magnetized, and the jet ejected from the central engine is also highly magnetized. When these highly magnetized jet shells collide with each other, the magnetic field (MF) lines will be twisted to cause magnetic reconnection events, and the synchrotron radiation can be generated by the accelerated electrons in the reconnection region. Although the physical processes in generating the gamma-ray photons of these three popular models are different, all can explain the typical GRB energy spectra.

The polarization study of GRB prompt emission can shed light on important issues such as the MF configuration and radiation mechanism of the GRBs \citep{Lan2016,Lundman2018,Wang2023,Lan2020,Lan2021,Parsotan2022,Guan2023}. \cite{Toma2009} and \cite{Guan2023} studied the statistical properties of the time- and energy-integrated polarizations in GRB prompt phase. In \cite{Guan2023}, they found that current time-integrated polarization observations in GRB prompt phase could be interpreted by the synchrotron emission in an ordered MF. The time-resolved and energy-resolved polarizations had been predicted under the framework of the magnetic reconnection model \citep{Lan2020} and the internal shock model \citep{Lan2021}. However, none of these studies involve the rotation of the polarization angle (PA).

Recently, \cite{2019A&A...627A.105B} divided the prompt phase of GRB 170114A into nine time bins. Its PA changes by approximately -$90^{\circ}$ between the second and third time bins, and by approximately $90^{\circ}$ between the fifth and sixth time bins. \cite{Kole2020} divided the main burst of GRB 170101A into two time bins, and the PA also changes by approximately $90^{\circ}$ between the two time bins. In our first paper \citep{Wang2023}, we showed that the rotation of the PA is usually happen for slightly off-axis observations. However, the influences of the parameters on the PA rotation have not been studied.

In this paper, we use the polarization model proposed by \cite{Lan2020} to study the time-resolved polarization properties, especially the influences of the parameters on the PA rotations. This paper is organized as follows. In Section 2, we briefly introduce the polarization model and provide our numerical results. The conclusions and discussion are presented in Section 3.

\section{The Model and the numerical results}\label{emodels}

\subsection{The model}\label{models}

A simple physical picture of the radiation region is assumed here, as that in \cite{Uhm2015,Uhm2016} and \cite{Uhm2018}. A thin relativistic jet shell expands radially outward in space. The magnetic reconnection process in jet would simultaneously accelerate the electrons and the jet itself. The accelerated electrons in the magnetic field would emit synchrotron photons uniformly from all positions within the jet shell. In the comoving frame of the shell, the emission generated at each position is assumed to be isotropic. This jet shell with a top-hat structure starts to emit at the radius $r_{on}$ and stops at $r_{off}$.

The polarization model we use here is same as \cite{Lan2020,sl2024}, and the detailed calculation formula can be found there. Since PA of the top-hat jet with a toroidal MF or a axisymmetric random MF could only change abruptly by $90^{\circ}$ or stay as a constant, while it could evolve gradually for an aligned field, the MF configuration considered here is a large-scale ordered aligned MF, which would be in the wind from a perpendicular rotator (i.e., the pulsar with its rotational axis perpendicular to its magnetic axis \citep{Spruit2001}). The time-resolved PD ($PD$) and preliminary PA ($PA_{pre}$) of the radiation from a jet with an aligned field in its emission region can be expressed as follows.
\begin{equation}
  PD=\frac{\sqrt{Q_\nu^2+U_\nu^2}}{F_\nu},
\end{equation}
\begin{equation}\label{PA}
  PA_{pre}=\frac{1}{2}\arctan(\frac{U_\nu}{Q_\nu}).
\end{equation}
where $F_\nu$ is the flux density, $Q_\nu$ and $U_\nu$ are the Stokes parameters Q and U, respectively. And the formula for these three Stokes parameters can be found in \cite{Lan2020,sl2024}. If $Q_\nu>0$, then the PA of the jet radiation is $PA=PA_{pre}$. If $Q_\nu<0$, then $PA=PA_{pre}+\pi/2$ for $U_\nu>0$ and $PA=PA_{pre}-\pi/2$ for $U_\nu<0$ \citep{Lan2018}.

One difference from \cite{Lan2020} is that the local PD used here is a broken power law.
\begin{equation}
    \Pi_{p,b} = \left\{
      \begin{array}{ll}
        (-\alpha_{B})/(-\alpha_{B}+\frac{2}{3}), & x\leq\alpha_{B}-\beta_{B},\\
        (-\beta_{B})/(-\beta_{B}+\frac{2}{3}), & x\geq\alpha_{B}-\beta_{B},
      \end{array}
    \right.
  \end{equation}
Here, the off-axis observations are also considered, the equal arrival time surface should include the off-axis observational geometry, which had already been done in \cite{Wang2023}.
\begin{equation}\label{eq:EATS}
t_{obs} = [t-\frac{r}{c}\cos\theta-t_{on}+\frac{r_{on}}{c}\cos\theta_0](1+z).
\end{equation}
where $\theta_0$ equals to 0 for on-axis observations, while it is $\theta_V-\theta_j$ for off-axis observations.

In the comoving frame, mainly due to the expansion of the jet shell, the MF strength in the radiation region decays with radius \citep{Uhm2014,Drenkhahn2002,Lan2021}
\begin{equation}
    B^{\prime}(r)=B_{0}^{\prime}(r/r_{0})^{-b}.
\end{equation}
Since the decay index of the MF ($b$) is unimportant for PA rotation \citep{Wang2023}, here we only consider two models with $b=1$ (i.e., $[2b_{i}]$ and $[2b_{m}]$) in \cite{Uhm2018}. The two models are parameterized, and the only difference between the ``i'' and ``m'' models is that their electron Lorentz factor $\gamma_{ch}$ varies differently with radius. The variation of $\gamma_{ch}$ in the $[2b_i]$ model is a single power law, and the corresponding peak-energy evolution mode is hard-to-soft.
\begin{equation}\label{gchi}
    \gamma_{ch}(r)=\gamma_{ch}^{0}(r/r_{0})^{g},
\end{equation}
where $\gamma_{ch}^{0}=5\times10^{4}$, and $g=-0.2$ \citep{Uhm2018}. While it is a broken power law for the ``m'' model, and the peak-energy evolution mode is intensity-tracking.
\begin{equation}\label{gchm}
    \gamma_{ch}(r)= \gamma_{ch}^{m}\times\left\{
      \begin{array}{ll}
        (r/r_{m})^{g}, & r\leq r_{m},\\
        (r/r_{m})^{-g}, & r\geq r_{m},
      \end{array}
    \right.
  \end{equation}
where we take $\gamma_{ch}^{m}=2\times10^{5}$, and $g=1.0$ \citep{Uhm2018}.

The variation of the bulk Lorentz factor $\Gamma$ is also roughly a power law with radius, as predicted in \cite{Drenkhahn2002}.
\begin{equation}
    \Gamma(r)=\Gamma_{0}(r/r_{0})^{s}.
\end{equation}
For an aligned field in the radiation region, $s$ equals to 0.35 \citep{Drenkhahn2002}.

\subsection{Numerical results}\label{Numerical}

With the model described in Section 2.1, we numerically calculate the detailed PA evolutions for different parameters. We define $\triangle$PA to be $\triangle$PA$\equiv$P$\textrm{A}_{max}-$P$\textrm{A}_{min}$, where P$\textrm{A}_{max}$ and P$\textrm{A}_{min}$ are the maximum and minimum values of the time-resolved PA within $T_{90}$, respectively. $T_{90}$ is the duration of time with the accumulated flux accounting from 5$\%$ to 95$\%$ of the total flux. And we only record the PA rotation with $\triangle$PA greater than $10^{\circ}$. The parameters we use are as follows: $\alpha_{B}=-0.8$, $\beta_{B}=-2.3$, $s=0.35$, $r_{on}=10^{14}$ cm, $r_{off}=3\times10^{16}$ cm, $r_{0}=10^{15}$ cm, $r_{m}=2\times10^{15}$ cm, $B_{0}^{\prime}=30$ G, $R_{inj}=10^{47}$ ${s}^{-1}$, and $\delta=\pi/6$ \citep{Uhm2018}. $R_{inj}$ is the injection rate of electrons in the shell, $\delta$ is the direction of the aligned MF, and $\theta_{j}$ is half-opening angle of the jet.

If the jet is accelerated thermally as $\Gamma$ is proportional to $r^1$, the later energy dissipation in the jet is very likely due to the internal shocks. In this scenario, the acceleration radius could be smaller than the radiation radius of GRB prompt phase. The thermal energy translates to the bulk kinetic energy of the jet first and then the bulk kinetic energy is dissipated at a larger radius through the shocks. However, in the scenario of the magnetic reconnection model, the jet is dominated by Poynting flux. Initially, the 
magnetic Reynolds number of the jet might be larger than 1, so the magnetic field is frozen in the jet and magnetic reconnection 
process can not happen. With the increase of the jet radius, its volume will increase, leading to a decrease of the particle number density. The conductivity of the jet, which is proportional to the particle number density, would also decrease. Hence its magnetic Reynolds would also decrease. Further more, the instabilities of the plasma would also develop during the jet propagation. Finally, at large radius the magnetic reconnection happens and the free magnetic energy is depleted. Actually, the energy for radiation and bulk acceleration of the jet in the magnetic reconnection model are both from the magnetic reconnection process, i.e., the acceleration radius is roughly same as the radiation radius in the scenario of the magnetic reconnection model. Since both the interpretations of the spectral lags in GRB prompt phase \citep{Uhm2016} and of the high-latitude emission in GRB prompt phase \citep{Li2021} suggested that the radiation region of GRB prompt emission is at large radius from the central engine ($10^{15}$–$10^{16}$ cm) and undergos the bulk acceleration, here we assume the radiation or the acceleration radii could be around $10^{15}$ cm.

It is worth noting that PA defined in the Section \ref{models} is within the range of ($-90^{\circ}$, $90^{\circ}$), which would lead to some abrupt $180^\circ$ jumps in the PA curves. Since the abrupt $180^\circ$ PA jump is meaningless and the polarization direction is unchanged, in this case we will first eliminate these abrupt $180^\circ$ jumps by adding or subtracting $180^\circ$ to keep the continuity of the PA curves at these jump points. Then $\triangle$PA$\equiv PA_{max}-PA_{min}$ is calculated with these handled PA curves. After this treatment, the maximum $\triangle$PA during the burst duration is $90^\circ$.

\begin{figure*}[htb]
    \centering
    \includegraphics[width=1\textwidth]{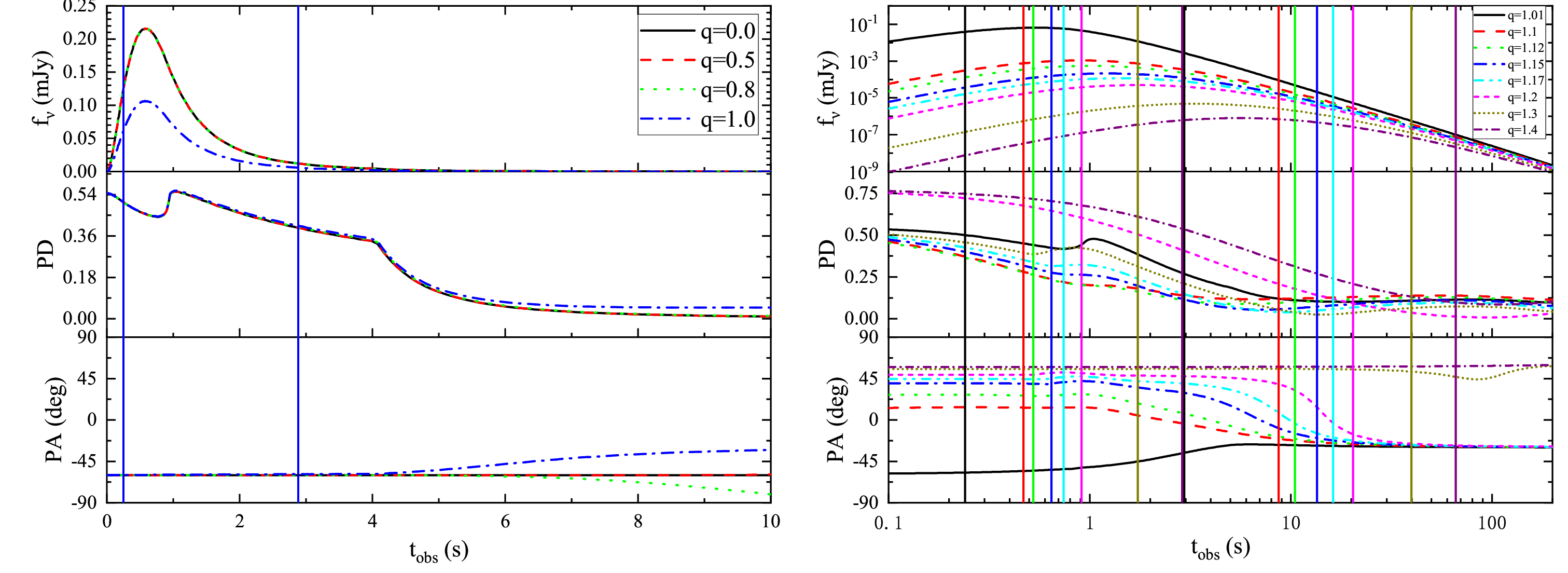}
    \caption{Light curves and polarization evolutions of the $[2b_{i}]$ model. The observational energies here is 300 keV. The top, middle, and bottom panels show the light curves, PD curves, and PA curves, respectively. On the left, the black solid, red dashed, green dotted and blue dash-dotted lines correspond to $q=0.0$, $0.5$, $0.8$ and $1.0$, respectively. On the right, the black solid, red dashed, green dotted, blue dash-dotted, cyan double dot-dashed, magenta short dashed, brown short-dotted, and purple short dash-dotted lines correspond to $q=1.01$, $1.1$, $1.12$, $1.15$, $1.17$, $1.2$, $1.3$, and $1.4$, respectively. The $T_{5}$ and $T_{95}$ of the light curve are shown as the vertical solid lines with the same color as the  corresponding light curve.
    }
    \label{tu1}
\end{figure*}

Figure \ref{tu1} shows the calculation results for the [$2b_{i}$] model. Its $\gamma_{ch}$ varies as a single power-law with radius (see Equation \ref{gchi}) and the corresponding peak-energy evolution mode is hard-to-soft. Here, we take $\Gamma_{0}=250$ and $\theta_{j}=0.1$ rad as typical parameters. In the first column, the observational angle $\theta_{V}$ is less than or equal to $\theta_{j}$, while in the second column, $\theta_{V}$ is slightly greater than $\theta_{j}$. We find that for on-axis observations(i.e.,$q\equiv\frac{\theta_{V}}{\theta_{j}}\leq1$), the PA remains unchanged within $T_{90}$. For off-axis observations (i.e.,$q>1$), $\triangle$PA$>10^{\circ}$ is within the range of $1.01\leq q\leq1.2$, and when $q>1.3$, the PA remains unchanged within $T_{90}$.

\begin{figure*}[htb]
    \centering
    \includegraphics[width=1\textwidth]{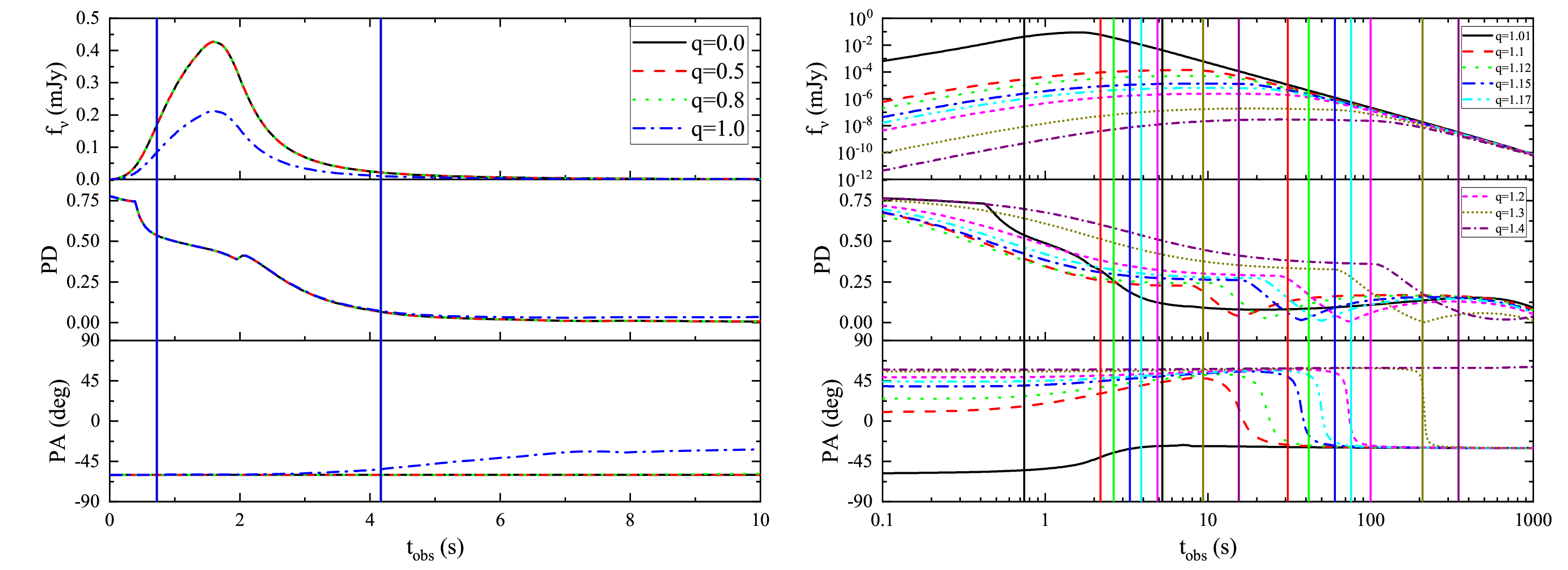}
    \caption{
        Same as Figure \ref{tu1}, but for model $[2b_{m}]$.
    }
    \label{tu2}
\end{figure*}

Figure \ref{tu2} shows the results for the [$2b_{m}$] model. The only difference between the [$2b_{i}$] and [$2b_{m}$] models is their variation of $\gamma_{ch}$ with radius. The $\gamma_{ch}$ of the $[2b_m]$ model varies as a broken power-law (see Equation \ref{gchm}) and the evolution mode of the peak energy is intensity-tracking. And also we take $\Gamma_{0}=250$ and $\theta_{j}=0.1$ rad. We find that the evolution of $\gamma_{ch}$ has almost no effect on the $q$ ranges for $\triangle$PA$>10^{\circ}$. For example, for the [$2b_{m}$] model, $\triangle$PA$>10^{\circ}$ is within the range of $1.01\leq q\leq1.3$.

The influences of $B^{\prime}$, $r_{on}$ and $r_{off}$ on the PA rotations are also studied and we find that the changes of these parameters also have slight effect on the q ranges for $\triangle$PA$>10^{\circ}$. The values of both $\Gamma_{0}$ and $r_{0}$ will affect the values of the bulk Lorentz factor $\Gamma$ within range from $r_{on}$ to $r_{off}$. And one profile of $\Gamma$ would correspond to lots of the combinations of ($\Gamma_0,r_0$). The changes of $\Gamma$ due to the changes of $r_0$ could be mimicked by the changes of $\Gamma_0$. So we take $r_0=10^{15}$ cm and only let $\Gamma_0$ vary. Therefore, in the following we will use [$2b_{i}$] model as an example and fix the values of $B^{\prime}$, $r_{on}$, $r_{off}$ and $r_{0}$ to explore the influence of the other parameters on the PA rotations.

\begin{figure*}[htb]
    \centering
    \includegraphics[width=1.0\textwidth]{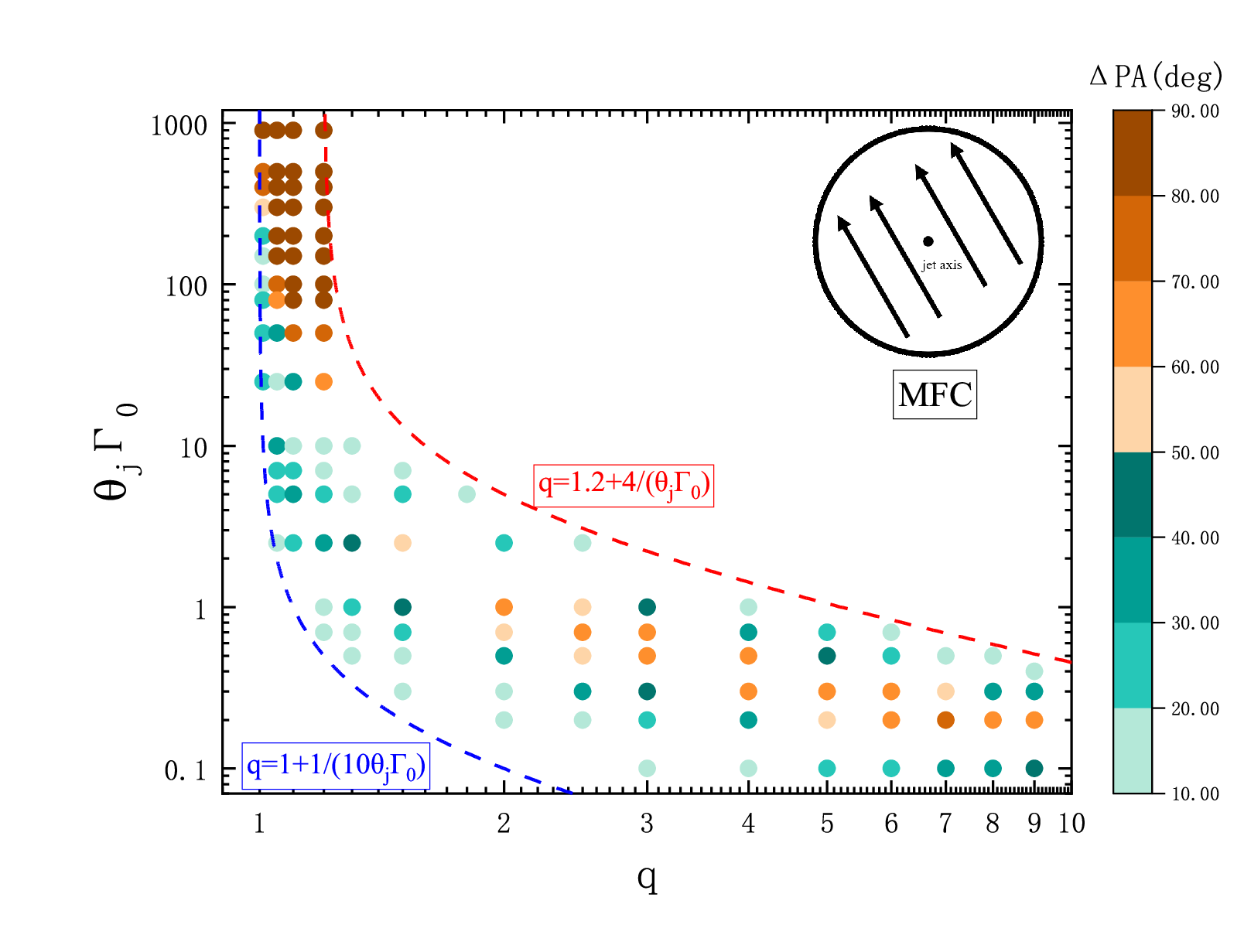}
    \caption{
        Distribution of the $\triangle$PA with $\theta_{j}\Gamma_{0}$ and $q$. The color changes from blue to black, indicating that $\triangle$PA will increase. The red-dashed and blue-dashed reference lines are the boundaries of the parameters with $\triangle$PA$>10^{\circ}$. In the upper right corner, a sketch diagram of the aligned magnetic field in the jet surface is shown.
    }
    \label{tu3}
\end{figure*}

We then calculated the PA rotations with different values of $\theta_{j}$, $\Gamma_{0}$ and $q$. A part of the detailed time-resolved results are shown in the Appendix. We found all three parameters have significant influence on the PA rotations within $T_{90}$. As shown in Figure \ref{tu3}, when the product value $\theta_{j}\Gamma_{0}$ is fixed, regardless of the concrete values of both $\theta_{j}$ and $\Gamma_{0}$, $\triangle$PA remains roughly unchanged for a fixed $q$ value. For example, for $\theta_{j}\Gamma_{0}=1$ with $q=2$, the values of $\triangle$PA are $66.99^{\circ}$ for $(\theta_{j},\Gamma_{0})=(0.02 rad,50)$, $65.88^{\circ}$ for $(\theta_{j},\Gamma_{0})=(0.01 rad,100)$, and $64.92^{\circ}$ for $(\theta_{j},\Gamma_{0})=(0.002 rad,500)$. Therefore, for fixed values of both $\theta_{j}\Gamma_{0}$ and $q$, $\triangle$ PA is roughly the same.

To demonstrate our above conclusion, we fix the product value of $\theta_{j}\Gamma_{0}$ and calculate the $\triangle$PA for various combinations of ($\theta_j,\Gamma_0$). The results are shown in Table \ref{table 1}. When $\theta_{j}\Gamma_{0}=0.1$, three sets of ($\theta_j,\Gamma_0$) are considered and the differences of the $\triangle$PAs between the various combinations are all within $10^\circ$ for one $q$ value. For $\theta_{j}\Gamma_{0}=1$, six sets of ($\theta_j,\Gamma_0$) are considered. The differences of $\triangle$PAs for most combinations are within $10^\circ$ except for $(\theta_{j},\Gamma_{0})=(0.001\rm{\ rad},1000)$ with $q=2.5$ and 3. Actually, the differences of the $\triangle$PAs between $(\theta_{j},\Gamma_{0})=(0.001 rad,1000)$ and other combinations for each $q$ value considered here are the maximum.

For $\theta_{j}\Gamma_{0}=10$, the differences of the $\triangle$PAs for the majority combinations are within $10^\circ$. However, they are larger than $10^\circ$ between $(\theta_{j},\Gamma_{0})=(0.01 rad,1000)$ and other combinations for $q=1.2$ and 1.3. When $q=1.3$, the differences of the $\triangle$PAs are relatively large and the maximum difference reaches $\sim11.8^\circ$. We only simply take $\triangle$PA$=19.92^\circ$ with parameter set of $(\theta_{j},\Gamma_{0})=(0.04 rad,250)$ (close to the typical set of $(\theta_{j},\Gamma_{0})=(0.1 rad,250)$) for $q=1.3$, as shown in Figure \ref{tu3}. For $\theta_{j}\Gamma_{0}=100$, the differences of the $\triangle$PAs between the six sets of ($\theta_j,\Gamma_0$) are all within $10^\circ$ for each $q$ value. Although there are some exceptions, the differences of the $\triangle$PAs will be within $10^\circ$ for the majority combinations of ($\theta_{j},\Gamma_{0}$) with the same values of both $\theta_{j}\Gamma_{0}$ and $q$,. Therefore, the conclusion is approximately held.

\renewcommand\arraystretch{1.2}
\setlength{\tabcolsep}{12pt}
\begin{table}[h]
\large
\caption{The values of $\triangle$PA within $T_{90}$ for different sets of ($\Gamma_0,\theta_j$) with a fixed product value of $\Gamma_0\theta_j$.}
\label{table 1}
\begin{tabular*}{\hsize}{l|cccccc}
\hline
\hline
$\Gamma_0\theta_j=0.1$ & $\Gamma_0=5$ & $\Gamma_0=10$ & $\Gamma_0=100$ & {} & {} & {}  \\\hline
$q=3$   & 11    &  11.2   & 11.66 & {} & {} & {}  \\
$q=4$   & 15.56 &  15.38  & 16.58 & {} & {} & {}  \\
$q=5$   & 19.88 &  20.42  & 21.19 & {} & {} & {}  \\
$q=6$   & 24.94 &  25.63  & 27.4  & {} & {} & {}  \\
$q=7$   & 31.31 &  33     & 34.01 & {} & {} & {}  \\
$q=8$   & 39.39 &  39.97  & 42.04 & {} & {} & {}  \\
$q=9$   & 48.02 &  49.3   & 51.21 & {} & {} & {}  \\\hline
\hline
$\Gamma_0\theta_j=1$ & $\Gamma_0=10$ & $\Gamma_0=50$ & $\Gamma_0=80$ & $\Gamma_0=100$ & $\Gamma_0=500$ & $\Gamma_0=1000$  \\\hline
$q=1.2$ & 17.45 &  17.38  & 17.23 & 17.58 &  20.58 &  23.24  \\
$q=1.3$ & 25.36 &  25.38  & 25.27 & 25.78 &  30.21 &  33.5   \\
$q=1.5$ & 40.55 &  40.4   & 40.41 & 40.83 &  45.74 &  48.81  \\
$q=2$   & 67.36 &  66.99  & 66.27 & 65.88 &  64.92 &  64.94  \\
$q=2.5$ & 62.13 &  60.44  & 59.75 & 58.94 &  52.07 &  48.42  \\
$q=3$   & 43.94 &  42.54  & 42.44 & 41.32 &  34.65 &  31.24  \\
$q=4$   & 19.41 &  18.78  & 18.87 & 18.74 &  15.83 &  14.54  \\\hline
\hline
$\Gamma_0\theta_j=10$ & $\Gamma_0=40$ & $\Gamma_0=50$ & $\Gamma_0=100$ & $\Gamma_0=250$ & $\Gamma_0=500$ & $\Gamma_0=1000$  \\\hline
$q=1.05$ & 30.56 &  30.65 & 30.35 & 30.15 &  29.85 &  28.83  \\
$q=1.1$  & 22.33 &  23.4  & 23.15 & 19.93 &  18.43 &  16.65  \\
$q=1.2$  & 14.25 &  14.19 & 14.02 & 18.68 &  21.23 &  25.57  \\
$q=1.3$  & 15.88 &  15.12 & 14.07 & 19.92 &  22.56 &  25.87  \\\hline
\hline 
$\Gamma_0\theta_j=100$ & $\Gamma_0=400$ & $\Gamma_0=500$ & $\Gamma_0=1000$ & $\Gamma_0=1600$ & $\Gamma_0=2000$ & $\Gamma_0=2500$  \\\hline
$q=1.01$ & 17.75 &  17.88 & 15.87 & 14.06 &  13.2  &  12.73  \\
$q=1.05$ & 74.13 &  74.63 & 76.2  & 77.22 &  77.64 &  78.1   \\
$q=1.1$  & 85.18 &  85.17 & 85.83 & 85.52 &  85.63 &  85.82  \\
$q=1.2$  & 87.97 &  87.14 & 85.77 & 85.27 &  84.48 &  84.88  \\\hline
\end{tabular*}
\tablecomments{For other parameters, refer to the first paragraph of Section \ref{Numerical}.}
\end{table}

In Figure \ref{tu3}, as the $\theta_{j}\Gamma_{0}$ value increases, the $q$ range for $\triangle$PA$>10^{\circ}$ becomes smaller. When $q$ range for $\triangle$PA$>10^{\circ}$ reduces to $1.0<q\leq1.2$, as $\theta_{j}\Gamma_{0}$ value increases, its range remains roughly unchanged and meanwhile the $\triangle$PA becomes larger. It is found that $\triangle$PA will reach roughly $90^{\circ}$ for $\theta_{j}\Gamma_{0}>50$ and $1<q\leq1.2$. For a fixed value of $\theta_{j}\Gamma_{0}$, the range of $q$ with $\triangle$PA$>10^{\circ}$ begins at $q=1+1/(10\theta_{j}\Gamma_{0})$ and ends at $q=1.2+4/(\theta_{j}\Gamma_{0})$. The $\triangle$PA increase and then decrease with $q$ for a fixed $\theta_{j}\Gamma_{0}$ value when $\theta_{j}\Gamma_{0}\leq10$, while it will increase with $q$ when $\theta_{j}\Gamma_{0}>10$. For a fixed value of $q$ with $q>1.2$, $\triangle$PA also increase and then decrease with $\theta_{j}\Gamma_{0}$. For a fixed value of $q$ with $q\leq1.2$, $\triangle$PA will roughly increase with $\theta_{j}\Gamma_{0}$.

\section{Conclusions and discussion}\label{conclusions}

In this paper, we use a magnetic reconnection model to study the rotation of the PA in the GRB prompt phase, especially the influences of the parameters. The MFC in the emitting region is assumed to be a large-scale ordered aligned MF. We find that the PA remains unchanged within $T_{90}$ for on-axis observations. The variation patterns of both $\gamma_{ch}$ and $B^{\prime}$ have slight impact on the rotation of the PA. $r_{on}$ and $r_{off}$ also have slight effect on the $q$ ranges for $\triangle$PA$>10^{\circ}$. Three key parameters ($\theta_{j}$, $\Gamma_{0}$, and $q$) are found and all have significant influences on PA rotation.  

With the calculation, we find that when the product of $\theta_{j}$ and $\Gamma_{0}$ is a fixed value, regardless of the concrete $\theta_{j}$ and $\Gamma_{0}$ values, $\triangle$PA remains roughly unchanged for a fixed $q$ value. Hence, for a fixed $\theta_{j}\Gamma_{0}$ value, regardless of $\theta_{j}$ and $\Gamma_{0}$ values, the $q$ ranges for $\triangle$PA$>10^{\circ}$ remain roughly unchanged. As the $\theta_{j}\Gamma_{0}$ value increases, the $q$ ranges for $\triangle$PA$>10^{\circ}$ become narrow. When $q$ range for $\triangle$PA$>10^{\circ}$ reduces to $1< q\leq1.2$, with the increase of $\theta_{j}\Gamma_{0}$ value, it remains roughly unchanged and the $\triangle$PA value becomes larger.

For an observational geometry (i.e., one $q$ value), the differences of the $\triangle$PAs for the majority combinations of ($\theta_j,\Gamma_0$) with a fixed value of $\theta_{j}\Gamma_{0}$ will be within $10^\circ$. Therefore, the conclusion that $\triangle$PA is independent of concrete combinations of ($\theta_j,\Gamma_0$) is approximately held. For a fixed value of $\theta_{j}\Gamma_{0}$, the range of $q$ with $\triangle$PA$>10^{\circ}$ begins at $q=1+1/(10\theta_{j}\Gamma_{0})$ and ends at $q=1.2+4/(\theta_{j}\Gamma_{0})$. For a fixed value of $\theta_{j}\Gamma_{0}$, the $\triangle$PA increase and then decrease with $q$ for $\theta_{j}\Gamma_{0}\leq10$, and it will increase with $q$ for $\theta_{j}\Gamma_{0}>10$. For a fixed value of $q$ with $q>1.2$, $\triangle$PA also increase and then decrease with the product value of $\theta_{j}\Gamma_{0}$. Most of the PA rotations with $\triangle$PA$\sim90^\circ$ are within the range of the key parameters: $1.0<q\leq1.2$ and $\theta_{j}\Gamma_{0}>50$.

All three key parameters found here ($\theta_{j}$, $\Gamma_{0}$ and $q$) are related to the geometries of the source and of the observation. With the calculation, the PA would rotate around the time when $\tilde{f}\sim1$ (except for the cases that $\tilde{f}$ is always 0.). The $\tilde{f}$ parameter was defined as the flux ratio between that from the region inside the $1/\Gamma$ cone and that from the outside \citep{Lan2020}. The time when $\tilde{f}\sim1$ corresponds to the transition time when the observed flux is dominated by the emission within the $1/\Gamma$ cone to that from the outside. Our results here are consistent with the conclusions made in \cite{Wang2023} that the changes of the observed images would lead to the PA rotation \citep{Wang2023} (see Figure 7 in their paper).

For current observations, only relatively bright bursts, which indicate on-axis or at least slightly off-axis observations could be analyzed with polarization. It is almost impossible for an aligned field in a top-hat jet that PA will rotate within $T_{90}$ in GRB prompt phase with on-axis observations. Therefore, if the jet structure of the GRBs is indeed the top-hat and the magnetic field is aligned, observations of the PA rotation within $T_{90}$ will indicate a slightly off-axis observations. Recent work suggested that the jet of GRBs may be structured \citep{Lan2023}. And how the PA rotates with the structured jet needs further studies. Same as in the afterglow phase \citep{Lan2016}, a gradual PA rotation in the GRB prompt phase would imply an aligned MFC in the emitting region, hence indicating a magnetar central engine \citep{Spruit2001}.

\begin{acknowledgements}
{We thank the anonymous referee for useful comments. This work is supported by the National Natural Science Foundation of China (grant No. 11903014). M.X.L also would like to appreciate the financial support from Jilin University.}
\end{acknowledgements}

\appendix
\restartappendixnumbering

\section{The Influences of The Key Parameters on the PA Rotation}\label{key}

Three key parameters that have significant influence on the PA rotation are the averaged bulk Lorentz factor $\Gamma_0$, the jet opening angle $\theta_j$, and the observational angle $\theta_V$. In the following calculations, the ranges of $q$ for $\triangle$PA$>10^{\circ}$ with different $\theta_{j}\Gamma_{0}$ values are studied. The fixed parameters we use are: $\alpha_{B}=-0.8$, $\beta_{B}=-2.3$, $s=0.35$, $r_{on}=10^{14}$ cm, $r_{off}=3\times10^{16}$ cm, $r_{0}=10^{15}$ cm, $B_{0}^{\prime}=30$ G, $R_{inj}=10^{47}$ s$^{-1}$ and $\delta=\pi/6$ \citep{Uhm2018}.

\begin{figure*}[htb]
    \centering
    \includegraphics[width=1\textwidth]{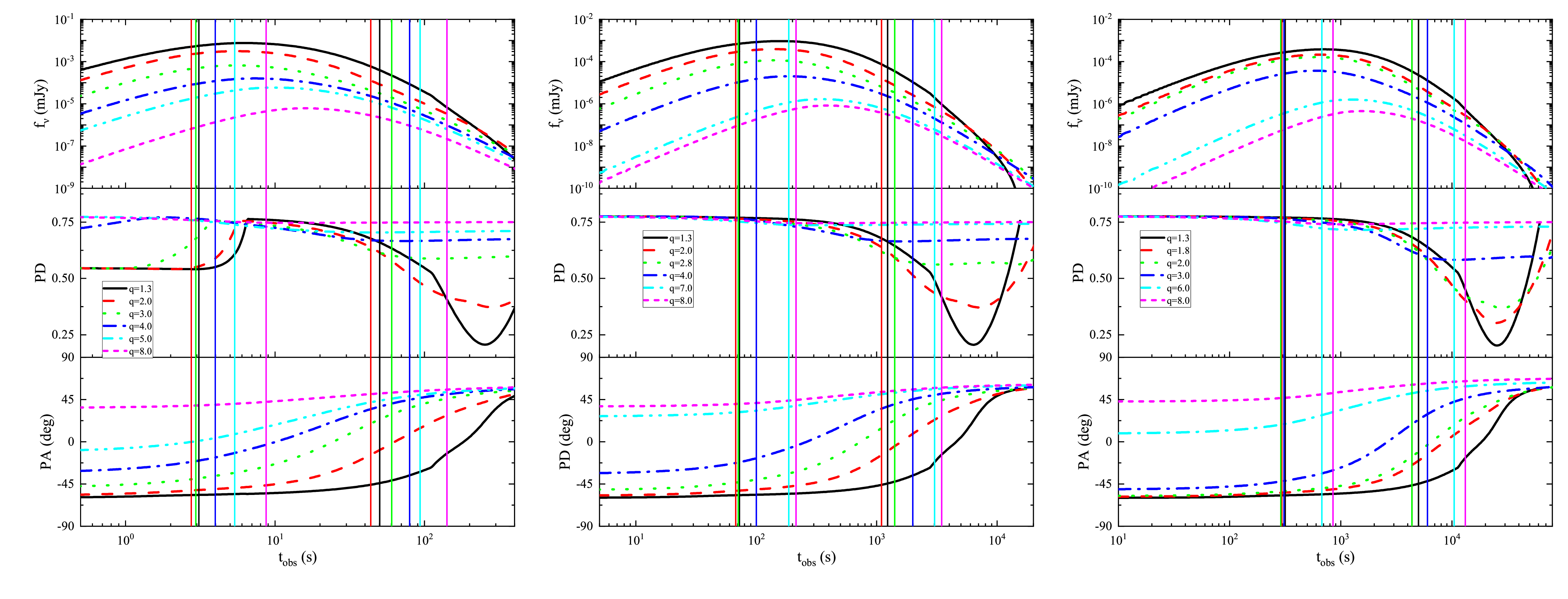}
    \caption{Light curves and polarization evolutions of the $[2b_{i}]$ model with $\theta_{j}\Gamma_{0}=0.5$. The observational energies here is 300 keV. The top, middle, and bottom panels show the light curves, PD curves, and PA curves, respectively. The ($\theta_{j}$, $\Gamma_{0}$) for the left, middle, and right panels are (0.01 rad, 50), (0.05 rad, 10), and (0.1 rad, 5), respectively. On the left panel, the black solid, red dashed, green dotted, blue dash-dotted, cyan double dot-dashed, and magenta short dashed lines correspond to $q=1.3$, $q=2.0$, $q=3.0$, $q=4.0$, $q=5.0$, and $q=8.0$, respectively. On the middle panel, the black solid, red dashed, green dotted, blue dash-dotted, cyan double dot-dashed, and magenta short dashed lines correspond to $q=1.3$, $q=2.0$, $q=2.8$, $q=4.0$, $q=7.0$ and $q=8.0$, respectively. On the right panel, the black solid, red dashed, green dotted, blue dash-dotted, cyan double dot-dashed, and magenta short dashed lines correspond to $q=1.3$, $q=1.8$, $q=2.0$, $q=3.0$, $q=6.0$, and $q=8.0$, respectively. The $T_{5}$ and $T_{95}$ of the light curve are shown as the vertical solid lines with the same color as the corresponding light curve.
    }
    \label{tuA1}
\end{figure*}

In Figure \ref{tuA1}, the value of $\theta_{j}\Gamma_{0}$ is set as 0.5. In the first column of Figure \ref{tuA1}, we take $\theta_{j}=0.01$ rad and $\Gamma_{0}=50$. When $1.3\leq q\leq8.0$, the value of $\triangle$PA is greater than $10^{\circ}$. Then we take $\theta_{j}=0.05$ rad and $\Gamma_{0}=10$, and the corresponding results are shown in the second column. When $1.3\leq q\leq8.0$, the value of $\triangle$PA is greater than $10^{\circ}$. Finally, we take $\theta_{j}=0.1$ rad and $\Gamma_{0}=5$, and the results are shown in the third column. When $1.3\leq q\leq8.0$, the value of $\triangle$PA is greater than $>10^{\circ}$. Through the above calculations, we find that for $\theta_{j}\Gamma_{0}=0.5$, the range of $q$ for $\triangle$PA$>10^{\circ}$ is approximately $1.3\leq q\leq8.0$.

\begin{figure*}[htb]
    \centering
    \includegraphics[width=1\textwidth]{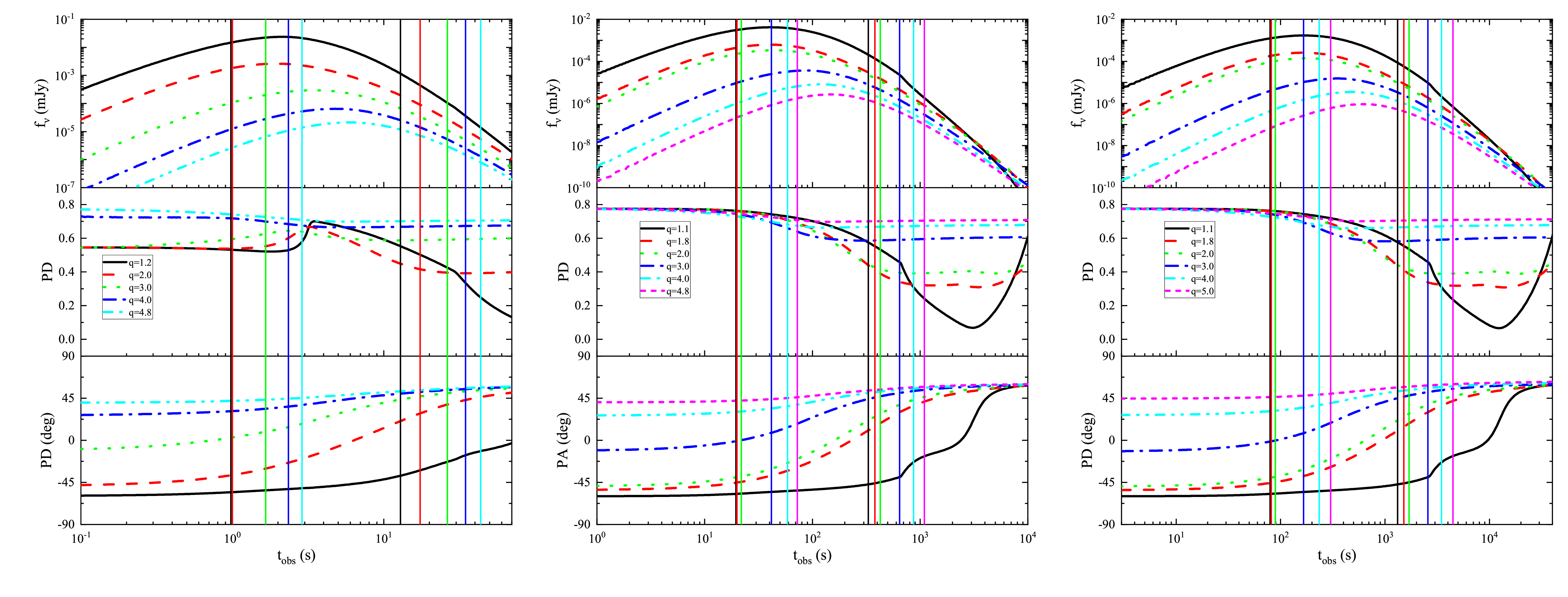}
    \caption{Same as Figure \ref{tuA1} but for $\theta_{j}\Gamma_{0}=1$. The left, middle and right panels correspond to $\Gamma_{0}=100$, $\Gamma_{0}=20$ and $\Gamma_{0}=10$, respectively. On the left panel, the black solid, red dashed, green dotted, blue dash-dotted, and cyan double dot-dashed lines correspond to $q=1.2$, $q=2.0$, $q=3.0$, $q=4.0$, and $q=4.8$, respectively. On the middle panel, the black solid, red dashed, green dotted, blue dash-dotted, cyan double dot-dashed, and magenta short dashed lines correspond to $q=1.1$, $q=1.8$, $q=2.0$, $q=3.0$, $q=4.0$ and $q=4.8$, respectively. On the right panel, the black solid, red dashed, green dotted, blue dash-dotted, cyan double dot-dashed, and magenta short dashed lines correspond to $q=1.1$, $q=1.8$, $q=2.0$, $q=3.0$, $q=4.0$, and $q=5.0$, respectively.
    }
    \label{tuA2}
\end{figure*}

In Figure \ref{tuA2}, we take $\theta_{j}\Gamma_{0}=1$, with $(\theta_{j},\Gamma_{0})=(0.01 rad,100)$ for column 1, $(\theta_{j},\Gamma_{0})=(0.05 rad,20)$ for column 2, and $(\theta_{j},\Gamma_{0})=(0.1 rad,10)$ for column 3. We find that for $\theta_{j}\Gamma_{0}=1$, the range of $q$ for $\triangle$PA$>10^{\circ}$ is approximately $1.2\leq q\leq4.8$.

\begin{figure*}[htb]
    \centering
    \includegraphics[width=1\textwidth]{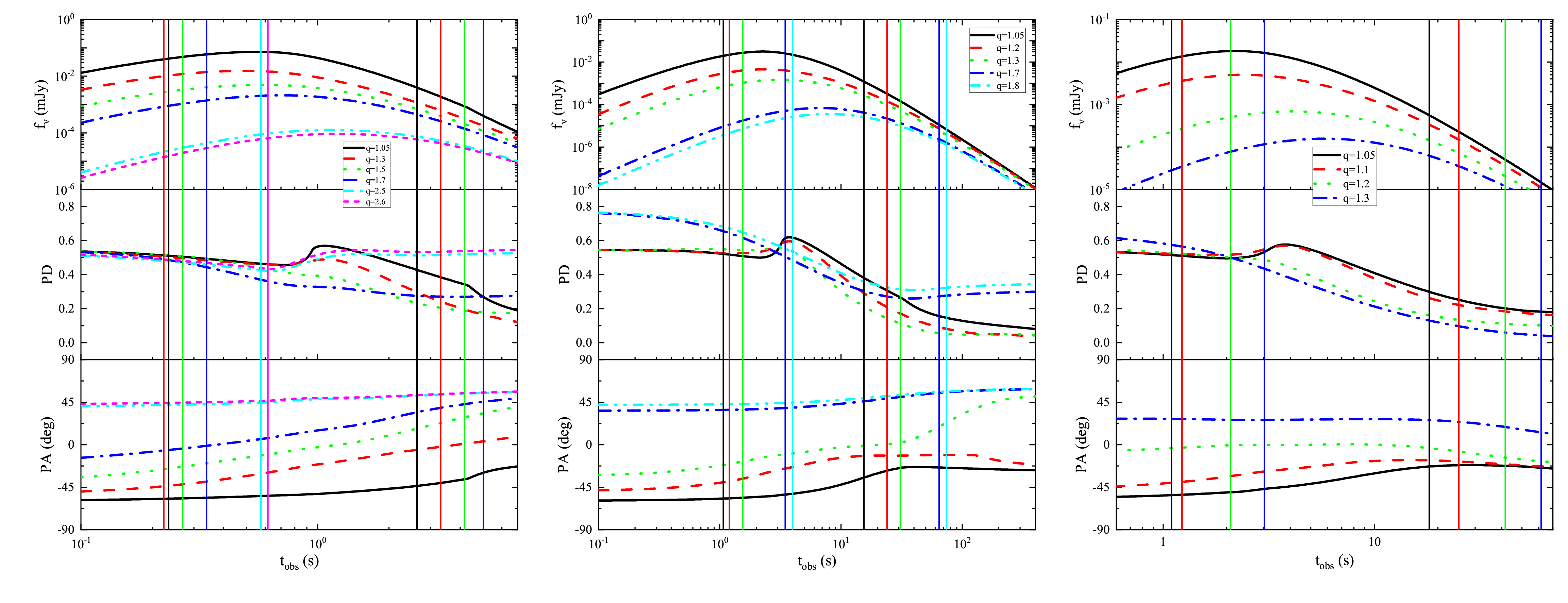}
    \caption{Same as Figure \ref{tuA1} but the left, middle and right panels correspond to $\theta_{j}\Gamma_{0}=2.5$, $5$ and $10$, respectively. The $\Gamma_{0}$ values for the left, middle and right panels are $250$, $100$, and $100$, respectively. On the left panel, the black solid, red dashed, green dotted, blue dash-dotted, cyan double dot-dashed, and magenta short dashed lines correspond to $q=1.05$, $q=1.3$, $q=1.5$, $q=1.7$, $q=2.5$, and $q=2.6$, respectively. On the middle panel, the black solid, red dashed, green dotted, blue dash-dotted, and cyan double dot-dashed lines correspond to $q=1.05$, $q=1.2$, $q=1.3$, $q=1.7$ and $q=1.8$, respectively. On the right panel, the black solid, red dashed, green dotted, and blue dash-dotted lines correspond to $q=1.05$, $q=1.1$, $q=1.2$, and $q=1.3$, respectively.
    }
    \label{tuA3}
\end{figure*}

In Figure \ref{tuA3}, we take $\theta_{j}\Gamma_{0}=2.5$ with $\theta_{j}=0.01$ rad and $\Gamma_{0}=250$ for the first column, $\theta_{j}\Gamma_{0}=5$ with $\theta_{j}=0.05$ rad and $\Gamma_{0}=100$ for the second column, and $\theta_{j}\Gamma_{0}=10$ with $\theta_{j}=0.1$ rad and $\Gamma_{0}=100$ for the third column. In the first column, $\triangle$PA$>10^{\circ}$ is within the range of $1.05\leq q\leq2.6$. In the second column, $\triangle$PA$>10^{\circ}$ is within the range of $1.05\leq q\leq1.8$. In the third column, $\triangle$PA$>10^{\circ}$ is within the range of $1.05\leq q\leq1.3$. So with the increases of $\theta_{j}\Gamma_{0}$ value, the $q$ ranges for $\triangle$PA$>10^{\circ}$ decreases.

\begin{figure*}[htb]
    \centering
    \includegraphics[width=1\textwidth]{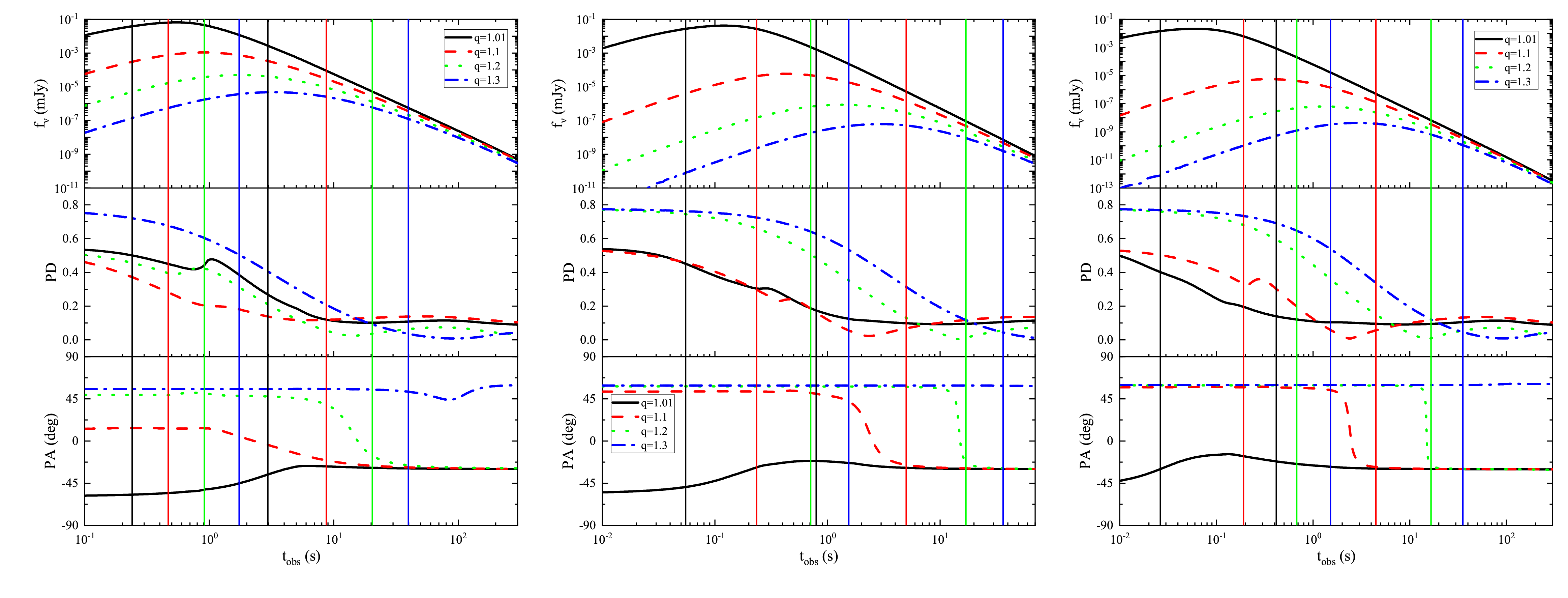}
    \caption{Same as Figure \ref{tuA1} but the $\theta_{j}\Gamma_{0}$ values for the left, middle, and right panels are $25$, $60$, and $100$, respectively. The half-opening angles of jet for the left, middle, and right panels are all 0.1 rad. The black solid, red dashed, green dotted, and blue dash-dotted lines correspond to $q=1.01$, $q=1.1$, $q=1.2$, and $q=1.3$, respectively. 
    }
    \label{tuA4}
\end{figure*}

In Figure \ref{tuA4}, we take $\theta_{j}\Gamma_{0}=25$ ($\theta_{j}=0.1$ rad and $\Gamma_{0}=250$) for column 1, $\theta_{j}\Gamma_{0}=60$ ($\theta_{j}=0.1$ rad and $\Gamma_{0}=600$) for column 2, and $\theta_{j}\Gamma_{0}=100$ ($\theta_{j}=0.1$ rad and $\Gamma_{0}=1000$) for column 3. We find that for $\theta_{j}\Gamma_{0}=25$, 60 and 100, the ranges of $q$ for $\triangle$PA$>10^{\circ}$ are all $1.01\leq q\leq1.2$, and the most significant PA rotation will happen when $\theta_{j}\Gamma_{0}\thicksim100$.

\listofchanges

\end{document}